\documentclass[prd,twocolumn,showpacs,superscriptaddress,nofootinbib,floatfix,showkeys,10pt]{revtex4-2}

\usepackage{graphicx}
\usepackage{amsmath}
\usepackage{bm}
\usepackage{yhmath}
\usepackage{mathtools}
\usepackage{wasysym}
\usepackage[colorlinks,citecolor=blue,urlcolor=blue,linkcolor=blue]{hyperref}
\usepackage{subfigure}
\usepackage{color}
\usepackage{cases}
\usepackage{subfigure}
\usepackage{times}
\usepackage{dcolumn,booktabs,bm}
\usepackage{slashed}
\usepackage{amsfonts,amssymb,stmaryrd,latexsym,amsmath}
\usepackage{textcomp}
\usepackage{multirow}
\usepackage{cancel}
\usepackage{array}
\usepackage{orcidlink}
\usepackage{enumitem}
\usepackage{dashrule}

\allowdisplaybreaks[1]

\newcommand{\etal}{\textit{et al}. }
\renewcommand{\arraystretch}{1.8}

\begin{document}


\title{Bound and Resonant States of Muonic Few-Body Coulomb Systems: Extended Stochastic Variational Approach}

\author{Liang-Zhen Wen\,\orcidlink{0009-0006-8266-5840}}
\email{wenlzh\_hep-th@stu.pku.edu.cn}
\affiliation{School of Physics, Peking University, Beijing 100871, China}

\author{Shi-Lin Zhu\,\orcidlink{0000-0002-4055-6906}}\email{zhusl@pku.edu.cn}
\affiliation{School of Physics and Center of High Energy Physics, Peking University, Beijing 100871, China}

\begin{abstract}
We compute the bound and resonant states of hydrogen-like muonic ions ($\mu\mu p$, $\mu\mu d$, $\mu\mu t$), three-body muonic molecular ions ($pp\mu$, $pd\mu$, $pt\mu$, $dd\mu$, $dt\mu$, $tt\mu$) and four-body double-muonic hydrogen molecules ($\mu\mu pp,\mu\mu dd,\mu\mu tt$) using an extended stochastic variational method combined with complex scaling.
The approach provides a unified treatment of bound and quasibound states and achieves an energy accuracy better than $0.1~\mathrm{eV}$ across all systems studied.
Complete spectra below the corresponding $n=2$ atomic thresholds are obtained, including several previously unresolved shallow resonances in both three- and four-body sectors.

\end{abstract}

\maketitle

\section{Introduction}~\label{sec:intro}

A muon is a lepton whose mass is about 207 times that of the electron.
As a consequence, muonic hydrogen-like systems are extremely compact, with characteristic length scales reduced by nearly two orders of magnitude compared with their electronic counterparts.
This spatial contraction amplifies bound-state QED and nuclear-structure contributions, including vacuum polarization, self-energy, and nuclear finite-size effects.
High-precision spectroscopy of the muonic hydrogen atom $p\mu^-$ has therefore enabled an exceptionally accurate determination of the proton charge radius, giving rise to the well-known proton radius puzzle~\cite{Pohl:2013yb}.

When more than one heavy lepton participates, muonic systems enter a regime where few-body correlations become essential.
For doubly muonic atoms such as $\mu^-\mu^-p$, reliable theoretical descriptions require a consistent treatment that combines few-body quantum mechanics with bound-state QED.
Previous studies have shown that strong correlations in these systems substantially enhance their overall sensitivity to QED and nuclear-structure effects compared with ordinary electronic atoms~\cite{Pachucki:1996zza,Jentschura:2005dli}.

Muonic few-body systems also play a central role in muon-catalyzed fusion ($\mu\mathrm{CF}$).
The strong spatial contraction increases nuclear overlap and motivated the early proposal of muonic intramolecular fusion (IMF), for example,
$dd\mu \;\rightarrow\; t + p + \mu$ ,
as suggested in Refs.\cite{frank1947hypothetical,Sakharov1948}.
Building on the same idea, Vesman proposed a resonant formation mechanism for the $dd\mu$ molecule\cite{vesman1967pis}:
\begin{equation}
d\mu(1S) + D_2 \longrightarrow [(dd\mu)dee],
\end{equation}
where the binding energy released by the muonic atom $d\mu(1S)$ is transferred into the rovibrational excitation of the molecular complex.

Because the interaction between an excited muonic atom $X\mu(n=2)$ ($X=p,d,t$) and the nucleus $X$ is dominated by a long-range dipole force, the resonance spectrum of muonic molecules forms a dense series of levels below the $n=2$ threshold~\cite{shimamura1989series}.
This dense accumulation strongly enhances molecular formation, and the corresponding resonant rates are expected to exceed those of the ordinary $dd\mu$ bound-state formation~\cite{Froelich:1995zz}.
The process can be viewed as an analogue of the Vesman mechanism:
\begin{equation}
d\mu(2S) + D_2 \longrightarrow [(dd\mu^{*})dee].
\end{equation}

To study these muonic few-body systems, we solve the Schrödinger equation using the complex scaling method~\cite{Aguilar:1971ve,Balslev:1971vb,Aoyama:2006hrz} combined with explicitly correlated Gaussians (ECGs)~\cite{Mitroy:2013eom}.
To obtain accurate near-threshold resonances, we further develop an improved basis-selection scheme based on the stochastic variational method (SVM) augmented by elements of the Gaussian expansion method (GEM)~\cite{Hiyama:2003cu}, referred to as the extended SVM (ESVM). 

The remainder of this paper is organized as follows: 
In Sec.~\ref{sec:framework}, we outline the theoretical framework, including the Hamiltonian, wave function construction, complex scaling method, and the approach for analyzing spatial structures.  Section \ref{sec:3body} presents our results for three-body bound and resonant states in the systems $\mu\mu p$, $\mu\mu d$, $\mu\mu t$, $pp\mu$, $dd\mu$, $tt\mu$, $pd\mu$, $pt\mu$, and $dt\mu$, for both S- and P-wave channels.
In Sec.~\ref{sec:4body}, we present the numerical results for the 4-body systems $\mu\mu pp,\mu\mu dd,\mu\mu tt$.   
Finally, in Sec.~\ref{sec:sum}, we provide a brief summary and discussion.  

\section{Theoretical framework}~\label{sec:framework}

\subsection{Hamiltonian}~\label{subsec:Hamiltonian}

The nonrelativistic Hamiltonian of the N-body QED system can be expressed as
\begin{align}\label{eq:Hamiltonian}
H=\sum_i^N\left(m_i+\frac{\boldsymbol{p}_i^2}{2m_i}\right)+\sum_{i<j=1}^n V_{i j}\,,
\end{align}
where $m_i$ and $\boldsymbol{p}_i$ denote the mass and momentum of particle $i$ and $V_{ij}$ reprents the coulomb interaction between particles i and j. 
The interaction potential is explicitly given by
\begin{align}\label{eq:Vij}
V_{ij} = \alpha\frac{Q_i Q_j}{r_{ij}},
\end{align}
where $Q_i$ is the charge of particle i, and
$r_{ij} = \lvert \boldsymbol{r}_i - \boldsymbol{r}_j \rvert$ denotes the interparticle separation.
In this work, we study three- and four-body systems composed of one or two protons, deuterons, or tritons together with one or two muons.
The particle masses and the fine-structure constant $\alpha$ are taken from CODATA~\cite{Mohr:2024kco}, listed in Table~\ref{tab:mass}. We also compute the bound-state energies of the hydrogen-like two-body subsystems (muonic hydrogen, deuterium, and tritium), and the results are listed in Table~\ref{tab:twobody}.

\begin{table}[htbp]
    \centering
    \caption{The masses of the nuclei ($p$, $d$, $t$) and of the muon $\mu$ (in MeV), together with the fine-structure constant $\alpha$.  
All values are taken from CODATA~\cite{Mohr:2024kco}}
    \label{tab:mass}
    \begin{tabular*}{\hsize}{@{}@{\extracolsep{\fill}}cccccccc@{}}
\hline\hline
&  $m_p$ & $m_d$ & $m_t$
& $m_{\mu}$ & $\alpha$ \\
\hline & 938.27209  & 1875.61294 & 2808.92113 & 105.65837  & $\frac{1}{137.036 00}$ \\
\hline\hline
    \end{tabular*}
\end{table}

\begin{table}[htbp]
\renewcommand{\arraystretch}{1.4}
\centering
\caption{\label{tab:twobody} The binding energies and rms radii of  muonic hydrogen, deuterium, and tritium are presented below. }
\begin{tabular*}{\hsize}{@{}@{\extracolsep{\fill}}lcccccc@{}}

\hline\hline
  & $E\,[\mathrm{eV}]$  & $r^{{rms}}\,[\mathrm{pm}]$ \\
 \hline 
 $p\mu(1S)$ & -2528.49 & 0.493\\
 $p\mu(2S)$ & -632.12 & 1.845\\
 $p\mu(3S)$ & -280.94 & 4.097\\
 $p\mu(4S)$ & -158.03 & 7.248\\
 \hline
 $d\mu(1S)$ & -2663.20 & 0.468\\
 $d\mu(2S)$ & -665.80 & 1.752\\
 $d\mu(3S)$ & -295.91 & 3.890\\
 $d\mu(4S)$ & -166.45 & 6.882\\
 \hline
 $t\mu(1S)$ & -2711.24 & 0.460\\
 $t\mu(2S)$ & -677.81 & 1.721\\
 $t\mu(3S)$ & -301.25 & 3.821\\
 $t\mu(4S)$ & -169.45 & 6.760\\
\hline \hline
\end{tabular*}
\end{table}

\subsection{Wave function construction}~\label{subsec:wavefunction}

The single-particle wave-function space is given by the direct product of the spatial component $\chi_{r}$ and the spin component $\chi_{s}$.
\begin{equation}\label{eq:Abasis}
\psi = \mathcal{A}\left( \chi_{r} \otimes \chi_{s} \right),
\end{equation}
where $\mathcal{A}$ denotes the appropriate (anti)symmetrization operator.
The particles in each hydrogen-like few-body system are labeled sequentially as $1,2,3$ (and $4$). For the $pp\mu$, $tt\mu$, $\mu\mu p$, $\mu\mu d$ and $\mu\mu t$ systems, the two identical fermions are antisymmetrized:
$$
\mathcal{A} = 1 - P_{12},
$$
where $P_{12}$ exchanges particles 1 and 2. For the $dd\mu$ system, the two deuterons are identical bosons, and the wave function is symmetrized,
$$
\mathcal{A} = 1 + P_{12}.
$$

For the four-body systems $\mu\mu pp$ and $\mu\mu tt$, both the two muons and the two nuclei are identical fermions, and the total antisymmetrization operator is
$$
\mathcal{A} = (1 - P_{12})(1 - P_{34}),
$$
where $P_{12}$ acts on the muons and $P_{34}$ on the protons or tritons. For the $\mu\mu dd$ system, the two muons are fermions while the two deuterons are bosons, and the operator becomes
$$
\mathcal{A} = (1 - P_{12})(1 + P_{34}).
$$

The total orbital angular momentum \(L\) is a good quantum number because the interaction is assumed to be purely Coulombic. In the absence of spin-dependent interactions, the spin of each subset of identical particles is conserved. For instance, in the three-body \(pp\mu\) system, the total spin of the two protons, \(S_{pp}\), is a good quantum number. Similarly, in the four-body \(\mu\mu pp\) system, both the muon-pair spin, \(S_{\mu\mu}\), and the proton-pair spin, \(S_{pp}\), remain good quantum numbers. Operationally, the primary role of the spin degrees of freedom in these systems is to enforce the correct (anti-)symmetrization properties of the spatial part of the total wave function.

For the spatial part of the wave function, we employ the explicitly correlated Gaussian (ECG) method~\cite{Mitroy:2013eom}. The wave function is expanded using the basis:
\begin{equation}\label{eq:basisSpace}
\psi(r)=\theta_{KLM}(\mathbf{v})\exp \left(-\sum_{i>j=1}^N \alpha_{i j}\left(\mathbf{r}_i-\mathbf{r}_j\right)^2\right),  
\end{equation}
where $\alpha_{i j}$ are adjustable nonlinear parameters. Higher orbital angular momenta $L$ are incorporated through the global vector representation (GVR)~\cite{suzuki1998new}. The function $\theta_{KLM}$ is constructed from the spherical harmonic $Y_{LM}$ built on a global vector $\mathbf{v}$ acting on a global vector $\mathbf{v}$ and a polynomial in $|\mathbf{v}|^{2}$:
$$
\theta_{KLM}(\mathbf{v})
= |\mathbf{v}|^{2K+L} Y_{LM}(\hat{\mathbf{v}}),
\qquad
\mathbf{v} = \sum_{i=1}^{N} u_i\mathbf{r}_i ,
$$
The coefficients $u_i$ specify different geometric configurations of the few-body system.
One linear combination corresponds to the center-of-mass motion. To eliminate this redundant degree of freedom, we impose
\begin{equation}\label{eq:ucondition}
 \sum_{i=1}^{N} u_i = 0.    
\end{equation}

The integer $K\ge 0$ controls the short-range behavior of the basis function and enhances its flexibility.

Since the ECGs contains only quadratic polynomials of $\mathbf{r}$ in its exponent, Eq.~\eqref{eq:basisSpace} can be written in a compact matrix form using a real symmetric matrix $A$,
\begin{equation}\label{eq:ECG}
\begin{aligned}
\psi_{KLM}(u,A,\mathbf{r})
&= \theta_{KLM}(u^{T}\mathbf{r}) \exp\left(-\frac{1}{2}\mathbf{r}^{T} A \mathbf{r}\right). \\
    A_{ii}&=0, \qquad(i=1,2,...,N)\\ 
    A_{ij}& = -\alpha_{ij}, \qquad(i\neq j)
\end{aligned}
\end{equation}

To isolate the intrinsic motion of the system, we employ the Jacobi coordinate set $(\mathbf{r}^{jac}_1,\mathbf{r}^{jac}_2,...,\mathbf{r}^{jac}_{N-1},\mathbf{r}_{CM})$ instead of the laboratory-frame coordinates
$(\mathbf{r}_{1},\mathbf{r}_{2},\ldots,\mathbf{r}_{N})$. The two coordinate sets are related by a linear transformation $U$,
\begin{equation}
\mathbf{r}^{jac}_i=\sum_{j=1}^N U_{i j} \mathbf{r}_j, \quad \mathbf{r}_i=\sum_{j=1}^N (U^{-1})_{i j} \mathbf{r}^{jac}_j, \quad i=1, \ldots, N .
\end{equation}
The explicit form of $U$ is
\begin{equation}
U=\left(\begin{array}{ccccc}
1 & -1 & 0 & \cdots & 0 \\
\frac{m_1}{M_2} & \frac{m_2}{M_2} & -1 & \cdots & 0 \\
\vdots & \vdots & \vdots & \ddots & \vdots \\
\frac{m_1}{M_{N-1}} & \frac{m_2}{M_{N-1}} & \frac{m_3}{M_{N-1}} & \cdots & -1 \\
\frac{m_1}{M_N} & \frac{m_2}{M_N} & \frac{m_3}{M_N} & \cdots & \frac{m_N}{M_N}
\end{array}\right) ,
\end{equation}
where $M_k = m_1 + m_2 + \cdots + m_k$.

In the Jacobi coordinate set, the matrix $A^{jac}$ and the vector $u^{jac}$ in the Jacobi coordinates are obtained from their laboratory-frame counterparts through
\begin{equation}
    A^{jac} =(U^{-1})^T A U^{-1},\qquad
    u^{jac} = (U^{-1})^T u.
\end{equation}

Because the original wave function in Eq.~\eqref{eq:basisSpace} is translationally invariant, only the first (N-1) Jacobi coordinates contribute to the intrinsic dynamics. Consequently, only the $(N-1)\times (N-1)$ block of $A^{\mathrm{jac}}$ and the $(N-1)$-component part of $u^{\mathrm{jac}}$ are non-vanishing.
In the following, we simply denote these reduced (N-1)-dimensional quantities by $\mathbf{r}$, A, and u.

Employing the GVR for the orbital component allows all required matrix elements to be evaluated analytically.
The wave function is generated from the auxiliary function g through
\begin{equation}
    \psi(\mathbf{r})
= \frac{1}{B_{KL}}\int d\mathbf{\hat{e}} Y_{LM}(\mathbf{\hat{e}}) (\frac{d^{2K+L}}{d\alpha^{2K+L}}g(\alpha,\mathbf{e},u,A,\mathbf{r}))_{\substack{\alpha=0 \\ |\mathbf{\hat{e}}|=1}}.
\end{equation}
where
\begin{equation}
\begin{aligned}
    g(\alpha,\mathbf{e},u,A,\mathbf{r})&= exp(-\frac{1}{2}\mathbf{r}^{T} A \mathbf{r}+ \alpha \mathbf{\hat{e}}\cdot u^{T}\mathbf{r})\\
    B_{KL}&=\frac{4\pi (2K+L)!}{2^KK!(2K+2L+1)!!},
\end{aligned}
\end{equation}
The same procedure applies to the matrix elements of an operator \(\hat{O}\):
\begin{equation}
\begin{aligned}
 & \left\langle \psi_{KLM}(u,A,\mathbf{r})\right| \hat{O}\left|\psi_{K'LM}(u',A',\mathbf{r}) \right\rangle \\
= & \frac{1}{B_{K L} B_{K^{\prime} L'}} \iint \mathrm{~d} \hat{\boldsymbol{e}} \mathrm{~d} \hat{\boldsymbol{e}}' Y_{L M}(\hat{\boldsymbol{e}})^* Y_{L M}\left(\hat{\boldsymbol{e}}'\right)\times \\
& \frac{\mathrm{d}^{\kappa+\kappa^{\prime}}}{\mathrm{d} \alpha^\kappa \mathrm{d} \alpha^{\prime \kappa^{\prime}}}\left\langle g(\alpha,\mathbf{e},u,A,\mathbf{r})\right| \hat{O}\left|g(\alpha',\mathbf{e}',u',A',\mathbf{r})\right\rangle_{\substack{\alpha=\alpha'=0 \\
|\hat{\mathbf{e}}|=|\hat{\mathbf{e}}'|=1}},
\end{aligned}
\end{equation}
where
$$
\kappa = 2K+L,\qquad
\kappa' = 2K' + L.
$$
Explicit formulas for the overlap and Hamiltonian matrix elements are given in Ref.~\cite{suzuki1998new}.

\subsection{Complex scaling method}~\label{subsec:method}

Few-body systems comprise various sub-channels, including break-up and rearrangement processes. Near-threshold bound and resonant states can be investigated by means of the complex scaling method (CSM). Within this framework, the energies and decay widths of resonant states in few-body systems are obtained directly through an analytic continuation of the Schrödinger equation~\cite{Aguilar:1971ve,Balslev:1971vb,Aoyama:2006hrz}. This continuation is implemented via a complex rotation of the coordinate $\boldsymbol{r}$ and momentum $\boldsymbol{p}$, expressed as
\begin{align}\label{eq:complexRotation}
U(\theta) \boldsymbol{r}=\boldsymbol{r} e^{i \theta}, \quad U(\theta) \boldsymbol{p}=\boldsymbol{p} e^{-i \theta}.
\end{align}
Under the rotation, the Hamiltonian in Eq. (\ref{eq:Hamiltonian}) becomes
\begin{equation}\label{eq:HamiltonianComplex}
H(\theta)=\sum_{i=1}^N\left(\frac{\mathbf{p}_i^2 e^{-2 i \theta}}{2 m_i}\right)+\sum_{i<j=1}^N V_{i j}\left(\mathbf{r}_{i j} e^{i \theta}\right).
\end{equation}
Meanwhile, for the resonant states with pole positions within the range of the rotated angle, their wave functions become normalizable by integration, thereby solvable through localized bases in the same way as bound states. As a result, solving the complex-scaled Schr\"odinger equation will simultaneously yield the eigenenergies of bound states and resonant states within the rotated angle.

A typical distribution eigen-energies in the complex energy plane is: Bound states appear on the negative real axis of the energy plane. Continuum states form straight lines (rotated cuts) emerging from the corresponding thresholds with $\operatorname{Arg}(E)=-2 \theta$. Resonant states with mass $M_R$ and width $\Gamma_R$ are located at $E_R=M_R-i \Gamma_R / 2$, and only those satisfying $\left|\operatorname{Arg}\left(E_R\right)\right|<2 \theta$ can be obtained in the CSM framework. The positions of the bound and resonant states remain unchanged with the variation of the rotation angle. The CSM is therefore an effective tool for distinguishing between scattering states and near-threshold bound states or resonances. More detailed discussions can be found in Refs.~\cite{Lin:2022wmj,Chen:2023eri,Chen:2023syh}.

In the present calculations, the numerical accuracy of the complex energies is at the level of $0.1~\mathrm{eV}$. Accordingly, we retain only digits consistent with this precision when reporting the results.

\subsection{Spatial structure}

The root-mean-square (rms) radius is a good physical quantity for reflecting the spatial structure of the hadron states. For near-threshold states, this is also an auxiliary criterion for determining whether it is a bound state or a scattering state.

The definition of the rms radius under CSM is 
\begin{equation}
r_{i j}^{{rms,C}} \equiv \operatorname{Re}\left[\sqrt{\frac{\left(\Psi(\theta)\left|r_{i j}^2 e^{2 i \theta}\right| \Psi(\theta)\right)}{\left(\Psi(\theta) \mid \Psi(\theta)\right)}}\right],
\end{equation}
where the $\Psi(\theta)$ is the obtained complex wave function of the state. The round bra-ket represents the so-called c-product~\cite{Romo:1968tcz} defined as
\begin{equation}
\left(\phi_n \mid \phi_m\right) \equiv \int \phi_n(\boldsymbol{r}) \phi_m(\boldsymbol{r}) {d}^3\boldsymbol{r},
\end{equation}
without taking complex conjugate of the bra-state. This procedure ensures the function inside the integral is analytic, thereby the expectation value of the physical quantity remains stable as the rotation angle changes. The rms radius calculated from the c-product is generally not real for the resonance. However, it is real for the bound state.

\subsection{Stochastic variational method and Gaussian expansion method}
The stochastic variational method (SVM) is an optimization scheme in which trial basis parameters are generated randomly and selected according to their variational performance.
Each basis-parameter set $(\alpha_{ij},u_i,,K)$ is sampled stochastically within physically motivated ranges.

Once the initial sampling is completed, we obtain a set of $N$ basis functions. The energy of the ground state or any other state of interest is then determined by solving the generalized eigenvalue problem
\begin{equation}
    H_{ij} \,C_j= S_{ij}\,C_j,  \qquad i,j \in 1,2,\dots, N
\end{equation}
where $H_{ij}$ and $S_{ij}$ are the Hamiltonian and overlap matrix elements evaluated with the chosen basis functions.

The next standard step of the SVM procedure is the so-called growth and refinement stage.
In the growth step, newly generated random trial bases are added one by one and evaluated.
The candidate that yields the largest improvement in the target-state energy is retained.
In the refinement step, one of the previously selected basis functions is replaced by a newly generated trial basis so that the total number of basis functions remains fixed. The replacement is accepted only if it further lowers the target-state energy.

SVM-optimized ECG bases provide highly accurate variational energies for bound systems.
However, selecting basis functions that lower the energy cannot be applied to resonant states. Because resonance poles exhibit more complicated cusp-like behavior rather than a monotonically decreasing trend in finite basis sets.

The Gaussian expansion method (GEM) is a specialized expansion scheme built on the ECG framework.
It is highly efficient for three- and four-body bound and resonant states.
Its efficiency comes from the ability to include physically important subchannels.
In particular, scattering configurations can be incorporated directly through suitable Jacobi coordinates.

\begin{figure}[htbp]
  \centering
  \includegraphics[width=0.47\textwidth]{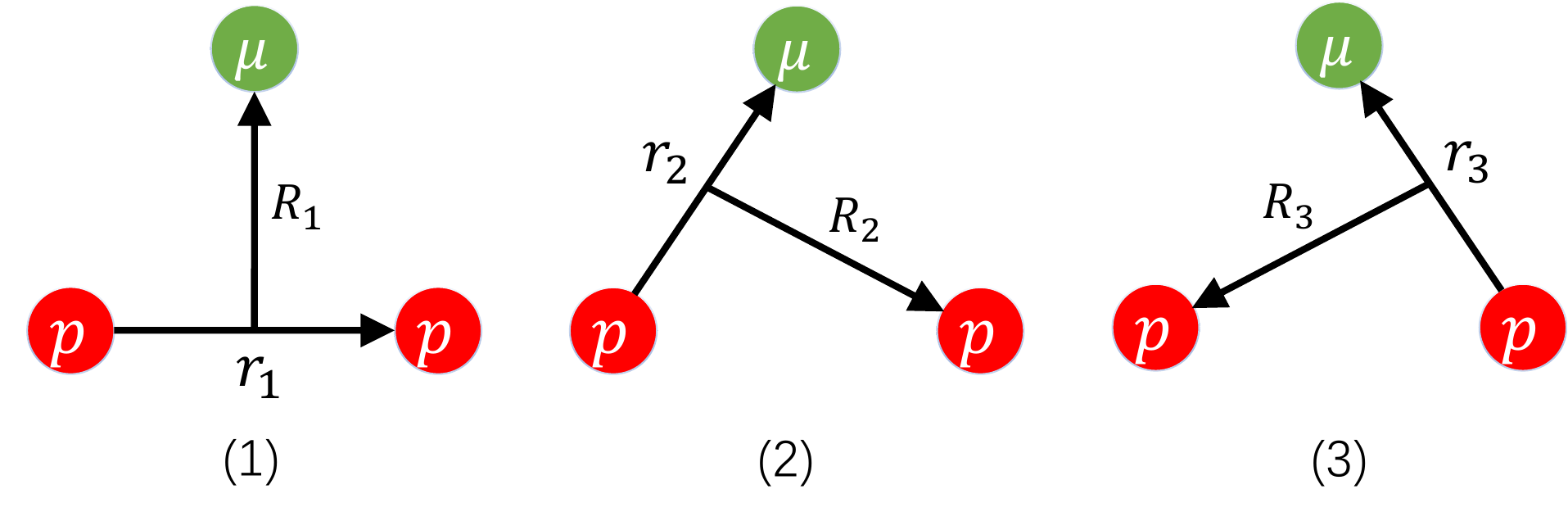} 
  \caption{\label{fig:structure} Jacobi coordinate configurations of the $pp\mu$ system in the conventional GEM framework. }
    \setlength{\belowdisplayskip}{1pt}
\end{figure}

In GEM, the spatial wave function is written as a direct product of single-Gaussian basis functions.
Each set of Jacobi coordinates defines one such product structure:
\begin{equation}\label{eq:Gaussians}
\begin{aligned}
\psi(\mathbf{r})&=\chi(\mathbf{r}_1)\chi(\mathbf{r}_2)\cdots \chi(\mathbf{r}_{N-1}),  \\
\chi_{n l m}(\boldsymbol{r})&=\sqrt{\frac{2^{l+5 / 2}}{\Gamma\left(l+\frac{3}{2}\right) r_n^3}}\left(\frac{r}{r_n}\right)^l e^{-\frac{r^2}{r_n^2}} Y_{l m}(\hat{r}),
\end{aligned}   
\end{equation}
where $r_n$ is taken as a geometric progression, 
$r_n=r_0a^{n-1}(n=1,2,..,n_{max}).$ For the three-body $pp\mu$ system,
the three Jacobi configurations used are shown in Fig.~\ref{fig:structure}.

 As we know, extending the method to systems with more particles faces dimensional difficulties:
\[
\begin{aligned}
\bullet\;& \text{For each Jacobi configuration, the basis size grows as } N^{\,n_{\max}}.\\[4pt]
\bullet\;& \text{The number of Jacobi configurations increases as } (2N-3)!!.
\end{aligned}\]

\subsection{Extended stochastic variational method}

We employ an extended stochastic variational method (ESVM), which consists of two stages.
The first stage follows the standard SVM procedure and is used to generate an initial set of basis functions.
The second stage introduces an extension in which scattering-like (molecular) configurations are incorporated in addition to the randomly generated ECG basis.

In the first stage, we generate an initial pool of 5000–10000 basis functions using the conventional SVM strategy.
The nonlinear parameters $\alpha_{ij}$ are sampled within physically motivated ranges.
For systems such as $\mu\mu pp$, the natural length scale is the Bohr radius of the muonic hydrogen atom, denoted by $r_{\mathrm{Bohr}}$.
To describe highly excited resonances, the basis must resolve both short-range and very long-range correlations.
Therefore, we sample $\alpha_{ij}$ logarithmically over an interval corresponding to $0.004$–$400$ Bohr radii of the relevant two-body subsystems:
\begin{equation}
\alpha_{ij} = \frac{1}{r_{ij}^{2}},  \qquad
r_{ij} \in [0.001,100] \,\mathrm{pm} .
\end{equation}

The coefficients $u_i$ are generated using a mixture of biased and unbiased Gaussian distributions, selected according to numerical tests and physical insight into the dominant subchannel geometries.

For the global-vector parameter $K$, we tested a uniformly sampled discrete distribution following the suggestion of E.~Mátyus~\cite{matyus2012molecular}. However, no improvement was found in the description of resonant states within our current numerical accuracy. We therefore set $K=0$ throughout this work.

\begin{figure}[htbp]
  \centering
  \includegraphics[width=0.47\textwidth]{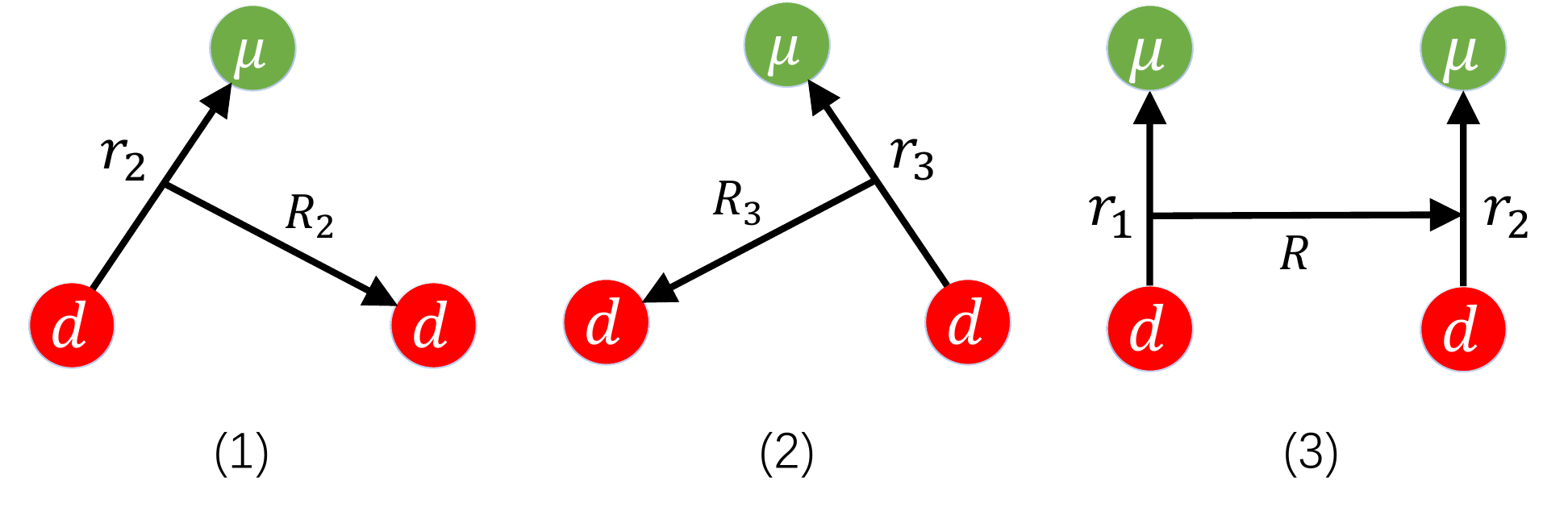} 
  \caption{\label{fig:scat} Molecular configurations for the $dd\mu$ and $\mu\mu dd$ systems in extended SVM. }
    \setlength{\belowdisplayskip}{1pt}
\end{figure}

The second stage forms the extension of the SVM.
In this stage, we add scattering-like basis functions.
We refer to these functions as molecular configurations.
These functions take the Gaussian form of Eq.~\eqref{eq:Gaussians}. They are constructed by combining sub-atomic clusters with relative functions between each pair of subclusters. For each molecular configuration, we use the Jacobi coordinate set that corresponds naturally to that configuration. Examples for the three- and four-body systems involving $d$ and $\mu$ are illustrated in Fig.\ref{fig:scat}. Because the two muons are identical fermions, only two distinct configurations are required for the $dd\mu$ system, while a single independent configuration is sufficient for the $\mu\mu dd$ system.
If the two nuclei are of the same type, the three-body configurations reduce to one independent structure, with the other obtained through the appropriate (anti-)symmetrization.

To construct these scattering-like basis functions, we first prepare optimized two-body wave functions.
Their parameters are chosen so that each subsystem reproduces the lowest energy of a selected target state.
To include highly near-threshold poles in few-body systems, we use the fourth excited atomic level as the optimization target.
For each subsystem, we generate 20 basis functions through 1000 refinement steps.
The sampling ranges are wide: the nonlinear parameters cover $0.004\text{–}400\, r_{\mathrm{Bohr}}$. The refinement procedure then selects the physically relevant values automatically.
The relative functions between subclusters are taken from a geometric progression. From this set, we keep 30–40 representative functions that span $0.1\text{–}80\, r_{\mathrm{Bohr}}$.
The corresponding range parameters satisfy
\begin{equation}
r_{0} = 0.02~\mathrm{\mathrm{pm}},
\qquad
r_{n_{\max}} = 20~\mathrm{\mathrm{pm}}.
\end{equation}

Instead of sampling the coefficients $u_i$ randomly, we prescribe them according to the molecular configuration of the relevant subclusters.
For instance, in a $p\mu(1S)$ atomic subsystem the intrinsic motion is purely S-wave; therefore, the component of the global vector along the Jacobi coordinate describing this subsystem is set to zero, $u_i = 0$.

\begin{figure}[htbp]
  \centering
  \includegraphics[width=0.5\textwidth]{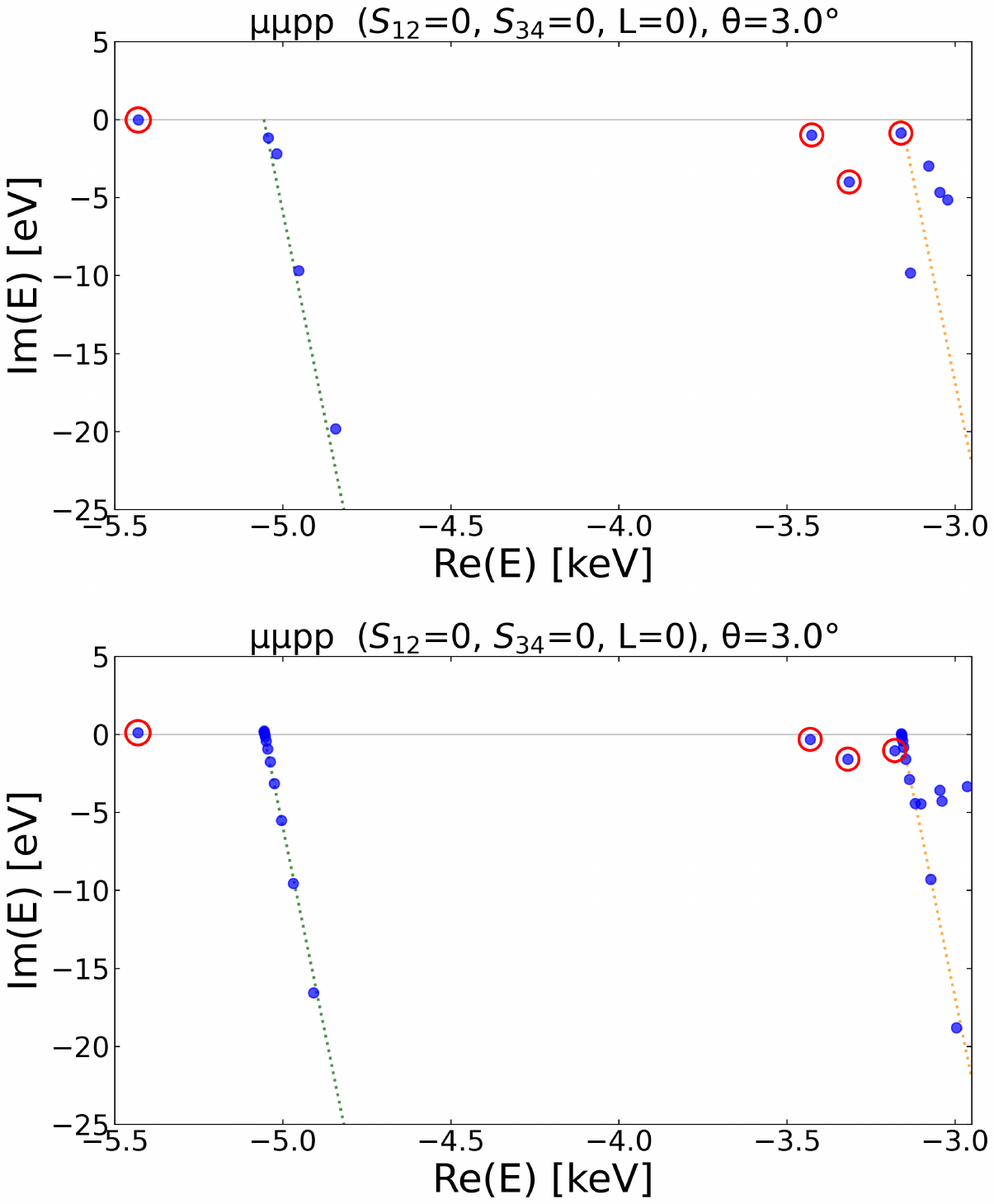} 
  \caption{\label{fig:comparison_scatter} Complex-energy spectra of the $\mu\mu pp$ system ($S_{12}=0$, $S_{34}=0$, $L=0$, $\theta=3^\circ$) obtained with SVM (top) and with ESVM (bottom). }
    \setlength{\belowdisplayskip}{1pt}
\end{figure}

Fig.~\ref{fig:comparison_scatter} shows the complex-energy spectra of the $\mu\mu pp$ system. A resonance just below the second threshold is hardly identifiable when only random bases are used (top), whereas it becomes clearly visible once scattering-like basis functions are included (bottom).
The extended SVM is particularly effective for poles lying near a specific scattering channel, because sizable components from that channel are expected due to the near degeneracy in energy.

\section{Three-body systems}~\label{sec:3body}

We employ the CSM to compute the complex eigenenergies of S- and P-wave three-body muonic Coulomb systems. The systems studied include
(A) hydrogen-atom–like muonic ions ($\mu\mu p$, $\mu\mu d$, $\mu\mu t$), and
(B) hydrogen-molecule–like muonic ions, comprising the $\mu$CF-related molecular ions ($dt\mu$, $dd\mu$, $tt\mu$) and the proton-containing systems ($pp\mu$, $pd\mu$, $pt\mu$).

Because of the structure of the basis functions, only states with natural parity are obtained,
$\text{Parity}=(-1)^L$ ,
where $L$ is the total orbital angular momentum.

For each system, we compute both S-wave and P-wave states, and we include both possible spin configurations of identical particle pairs ($S_{12}=0, 1$) whenever applicable.

The complex eigenenergies, spin compositions, and rms radii of the bound and resonant states are summarized in Tables~\ref{tab:mmX}–\ref{tab:ptm}. Rms values marked with an asterisk (*) indicate that the imaginary part is significant compared to the real part.
Benchmark energies from previous studies are also listed for comparison, demonstrating the accuracy and reliability of the present ESVM approach.

\subsection{hydrogen–like muonic ions }

\begin{figure}[htbp]
  \centering
  \includegraphics[width=0.45\textwidth]{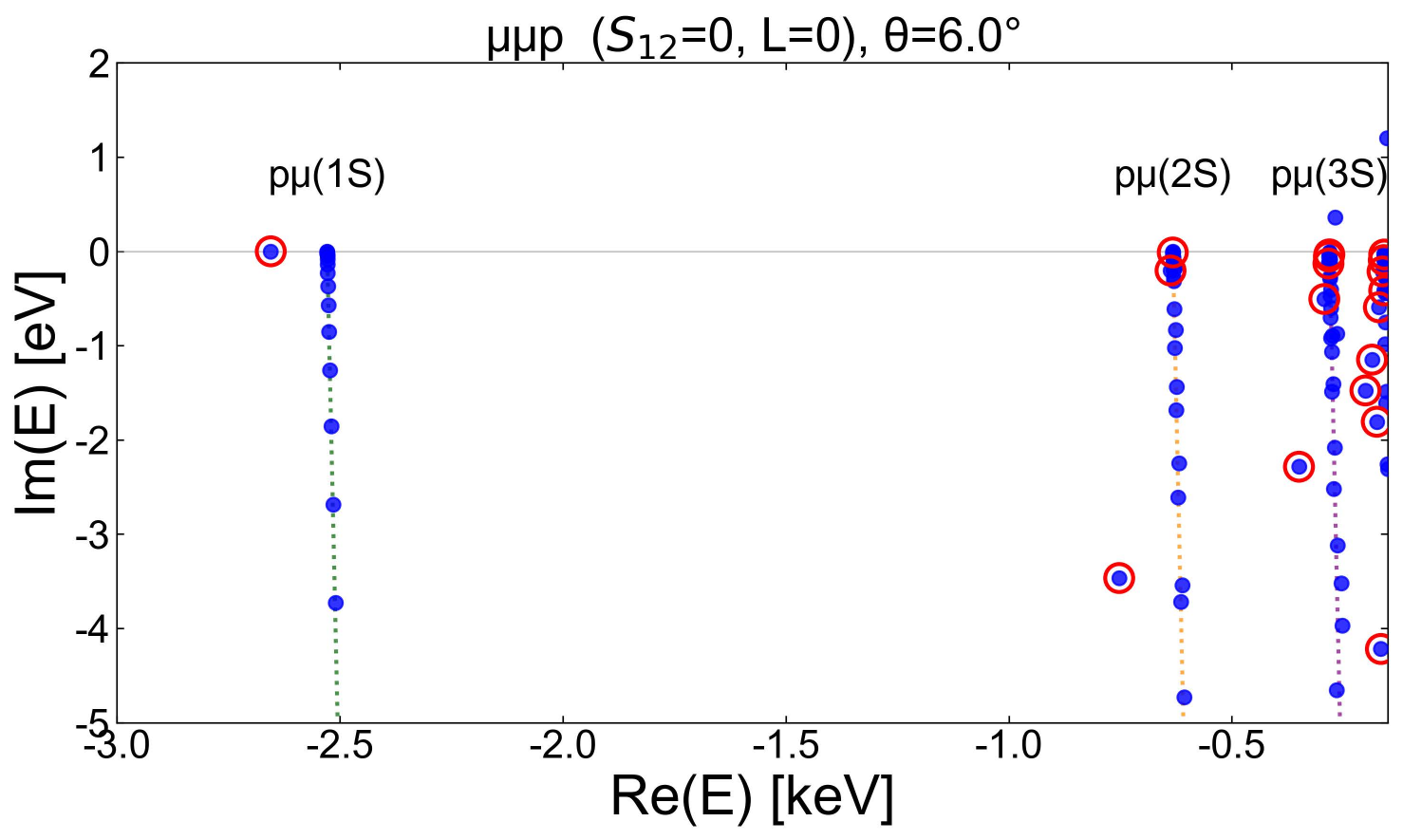} 
  \caption{\label{fig:mmp} Complex-energy spectra of the $\mu\mu p$ system ($S_{12}=0$, $L=0$, $\theta=6^\circ$) . }
    \setlength{\belowdisplayskip}{1pt}
\end{figure}

\begin{table}[htbp]
\renewcommand{\arraystretch}{1.0}
\centering
\caption{\label{tab:mmX}
Complex energies and rms distances for the $\mu\mu X$ systems ($X = p,d,t$).
$r^{\mathrm{rms}}_{\mu\mu}$ and $r^{\mathrm{rms}}_{\mu X}$ denote the muon–muon and muon–nucleus rms distances (pm), respectively. Benchmark energies from Refs.~\cite{liverts2013three,Yang:2025oih} are shown in the final column.
}

\begin{tabular*}{\hsize}{@{}@{\extracolsep{\fill}}lcccccc@{}}

\hline\hline
 system & $(S_{\mu\mu},L)$ & $E\,[\mathrm{eV}]$   & $r^{{rms}}_{{\mu\mu}}[\mathrm{pm}]$& $r^{{rms}}_{{\mu X}}[\mathrm{pm}]$ & $E_{bench}\,[\mathrm{eV}]$ &\\
 \hline 
 $\mu\mu p$
  & [0,0] & -2654.93  & 1.46 & 1.02 & 2654.93\\

 &    $\mathbf{p\mu(n=1)}$ & -2528.49  &  & \\
  & [0,0]& 752.54-3.5i  & 6.8 & 4.6 &-752.60-3.6i\\
  & [0,0]& -637.41-0.2i &  13.0  & 8.7     \\
  & [0,0]& -632.43 &  51.5  &  35.8    \\
  & [1,0]& -642.21 & 9.7   &  6.3    \\
  & [1,0]& -632.70 &  37.8  & 26.2     \\
  & [0,1]& -637.22 &  6.8  &  5.0    \\
  & [0,1]& -632.28&  32.2  &  22.9    \\
  & [1,1]& -720.48-0.1i &  2.2  &  1.8    \\
  & [1,1]& -634.29 &  10.1  &  7.3    \\

 &    $\mathbf{p\mu(n=2)}$  & -632.12 && & \\
 \hline
  $\mu\mu d$
  & [0,0]& -2802.55  & 1.38 & 0.95 &-2802.55\\
 & $\mathbf{d\mu(n=1)}$ & -2663.20    & \\

  & [0,0]& -792.31-4.3i  & 3.3 & 2.3 &-792.38-4.2i\\
  
  & [0,0]& -671.27-0.2i  & 12.6 &8.4\\
  & [0,0]& -666.11  & 49.5 &34.5\\
  & [1,0]& -676.66  & 9.1 & 5.9\\
  & [1,0]& -666.41  &35.7 & 24.7\\
  & [0,1]& -671.24  & 6.4 & 4.7\\
  & [0,1]& -665.97  & 31.2 & 22.2\\
  & [1,1]& -757.82-0.2i & 2.2 & 1.7\\
  & [1,1]& -668.01  & 9.8 & 7.0\\

 &  $\mathbf{d\mu(n=2)}$ & -665.80 && & \\
 \hline$\mu\mu t$
  & [0,0]& -2855.68  & 1.35 & 0.93 &-2855.68\\
 &   $\mathbf{t\mu(n=1)}$ & -2711.24& & & \\

  & [0,0]& -806.75-4.5i  & 2.9 & 1.7 & -806.65-4.4i\\
  & [0,0]& -683.35 - 0.2i & 12.3 & 8.2\\
  & [0,0]& -678.12 & 48.7 & 34.0\\
  & [1,0]& -688.97  & 8.9 & 5.8\\
  & [1,0]& -678.44  & 34.9 & 24.2\\
  & [0,1]& -683.39  & 6.2 &4.6\\
  & [0,1]& -677.98  & 30.7 & 21.8\\
  & [1,1]& -771.20 - 0.3i &2.1&1.7\\
  & [1,1]& -680.05  & 9.6 &6.9\\

 &    $\mathbf{t\mu(n=2)}$ &-677.81 & & &\\

\hline \hline
\end{tabular*}
\end{table}

\begin{figure}[htbp]
  \centering
  \includegraphics[width=0.45\textwidth]{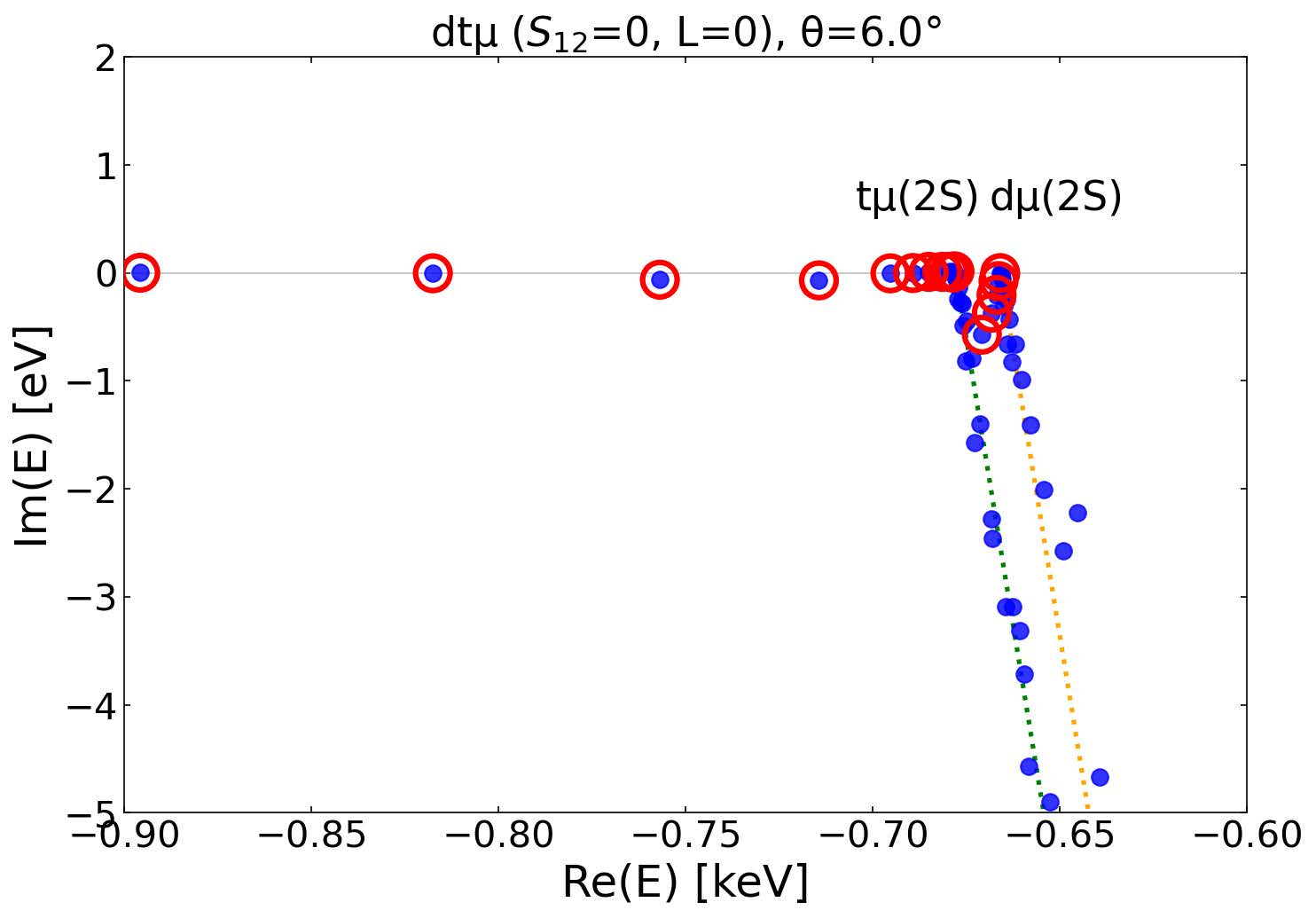} 
  \caption{\label{fig:dtm} Complex-energy spectra of the $dt \mu$ system ($S_{12}=0$, $L=0$, $\theta=6^\circ$) . }
    \setlength{\belowdisplayskip}{1pt}
\end{figure}

\begin{table}[htbp]
\setlength{\tabcolsep}{0.1pt} 
\renewcommand{\arraystretch}{1.0}
\centering
\caption{\label{tab:dtm} The binding energies and rms radii of  $dt\mu$ are presented below. Benchmark results from previous calculations are listed in the last column for comparison. Newly identified states are marked with a dagger ($\dagger$). }
\begin{tabular*}{\hsize}{@{}@{\extracolsep{\fill}}lcccccc@{}}

\hline\hline
  $L$ & $E\,[\mathrm{eV}]$   & $r^{{rms}}_{dt}[\mathrm{pm}]$ &$r^{{rms}}_{d\mu}[\mathrm{pm}]$&$r^{{rms}}_{t\mu}[\mathrm{pm}]$& $E_{bench}\,[\mathrm{eV}]$ \\
 \hline
   0& -3030.38  & 0.74 & 0.62 & 0.59 &-3030.38~\cite{kilic2004coulombic}\\
   0& -2746.08  & 1.42 & 1.21 & 0.88 &-2746.08~\cite{kilic2004coulombic}\\
   1 & -2943.71  & 0.41 & 0.49 & 0.47 &-2943.70~\cite{hara1989bound} \\
   1 & -2711.90  & 1.48 & 1.50 & 0.55& -2711.90~\cite{hara1989bound}\\

    $\mathbf{t\mu(n=1)}$ & -2711.24& &  \\

$\mathbf{d\mu(n=1)}$ & -2663.20  & &  & \\

   0& -895.70  & 2.6 & 1.8 & 1.7 \\
 0& -817.54  & 3.2 & 2.1 & 2.1 \\
   0& -756.92  & 3.9 & 2.6 & 2.4 \\
   0& -714.41  & 5.1 & 3.7 & 2.7 \\
   0& -695.26  & 6.9 & 5.8 & 2.7 \\
   0& -689.23  & 7.1 & 4.9 & 4.5 \\
   0& -685.04  & 10.2 & 9.2 & 2.9 \\
  0& -681.38  & 14.6 & 13.8 & 2.5 \\
   0& -679.54  & 21.0 & 20.2 & 2.4 \\
   0$^{\boldsymbol{\dagger}}$& -678.64  & 30.6 & 29.8 & 2.3 \\
   0$^{\boldsymbol{\dagger}}$& -678.20  & 48.3 & 47.6 & 2.0 \\

   1 & -890.35  & 1.3 & 1.3 & 1.3 \\
   1 & -813.19  & 1.6 &  1.5 & 1.5 \\
   1 & -753.52  & 2.0 &  1.8 & 1.6 \\ 
  1 & -711.96  & 2.6 &  2.6 & 1.3 \\
     1 & -696.97  & 3.2 &  2.3 & 2.0 \\
     1 & -694.11  & 3.5 &  3.3 & 1.8 \\
     1 & -688.30  & 3.9 &  3.4 & 2.3 \\
     1 & -684.23  & 5.6 &  5.4 & 1.6 \\
    1 & -680.97  & 8.0 &  7.7 & 1.7 \\
     1 & -679.32  & 11.6 &   11.4 & 1.8 \\
     1$^{\boldsymbol{\dagger}}$ & -678.52  & 17.2 &   16.9 & 1.8 \\
     1$^{\boldsymbol{\dagger}}$ & -678.14  & 23.6 &  23.3  &  1.6\\

  $\mathbf{t\mu(n=2)}$ &-677.81 & & &\\

   0& -670.73-0.6i  & 37.5 & 35.7 & 11.5 &-670.81-0.6i~\cite{kilic2004coulombic}\\
   0& -668.18 - 0.4i & $7.1^*$ & $7.3^*$ & 18.5 &-668.13-0.3i~\cite{kilic2004coulombic}\\
   0& -666.87 - 0.2i & 29.1 & 15.3 & 26.4 &-666.90-0.2i~\cite{kilic2004coulombic}\\
   0$^{\boldsymbol{\dagger}}$& -666.30 - 0.1i & 41.0 & 10.6 & 39.1 \\
   0$^{\boldsymbol{\dagger}}$& -666.03  & 60.8 & 6.9 & 59.7 \\
     1$^{\boldsymbol{\dagger}}$ & -670.62 - 0.7i & 23.7 &   23.1 & 5.9 \\
     1$^{\boldsymbol{\dagger}}$ & -668.03 - 0.4i & 7.9 &  $3.4^*$ & 9.1 \\
     1$^{\boldsymbol{\dagger}}$ & -666.82 - 0.2i & 13.0 &   $0.3^*$ & 13.5 \\
     1$^{\boldsymbol{\dagger}}$ & -666.25-0.1i  & 21.8 &   1.0 & 21.7 \\
1$^{\boldsymbol{\dagger}}$ & -665.97  & 35.6 &   2.4 & 35.3 \\

  $\mathbf{d\mu(n=2)}$ & -665.80 && & \\
\hline \hline
\end{tabular*}
\end{table}

The $\mu\mu X$ systems ($X = p, d, t$) are the muonic analogues of the negative hydrogen ion $\mathrm{H}^-$.
They consist of two negatively charged muons bound to a nucleus $X$ through the Coulomb interaction.
Because the muon is much heavier than the electron, nuclear recoils cannot be neglected and the full three-body dynamics must be treated explicitly.

Liverts and Barnea~\cite{liverts2013three} obtained high-precision results for the bound state and the lowest resonance using Laguerre-polynomial expansions.
Yang \etal\cite{Yang:2025oih} performed GEM calculations and reported results consistent with theirs.
Our fully dynamical calculations show excellent agreement with both sets of results, as summarized in Table~\ref{tab:mmX}.
For the three-body bound states of the $\mu\mu p$, $\mu\mu d$, and $\mu\mu t$ systems, the discrepancies remain below $0.01~\mathrm{eV}$.

Beyond the lowest states, the present ESVM approach allows us to compute all resonances up to the $X\mu(n=4)$ threshold, as shown in Fig.~\ref{fig:mmp}.
In Table~\ref{tab:mmX}, we summarize the results for states lying below the $X\mu(n=2)$ dissociation threshold.
The excited states exhibit larger spatial extent compared to the bound states.
Except for the ground bound state, all resonances possess complex rms radii due to their finite widths.
Since these widths are small, the imaginary parts of the rms radii are negligible, and the real parts provide useful structural information.

\subsection{muonic molecular ions}

The $dt\mu$, $dd\mu$ molecular ions are central objects of interest in $\mu$CF physics.
Their bound states and the resonances lying below the $n=2$ atomic thresholds have been investigated in great detail.
The shallow $S_{12}=1$, $L=1$ bound states in $dt\mu$ and $dd\mu$—which govern the resonant formation step in $\mu$CF—lie only $0.66~\mathrm{eV}$ and $1.97~\mathrm{eV}$ below the respective $n=1$ thresholds.
Our ECG calculations reproduce these benchmark energies within the quoted numerical accuracy, consistent with Refs.~\cite{kilic2004coulombic,Yamashita:2024inr} respectively.

\begin{table}[htbp]
\renewcommand{\arraystretch}{1.0}
\centering
\caption{\label{tab:ddm} The binding energies and rms radii of  $dd\mu$ and tritium are presented below. Newly identified states are marked with a dagger ($\dagger$). }
\begin{tabular*}{\hsize}{@{}@{\extracolsep{\fill}}lcccccc@{}}

\hline\hline
 $[S_{dd},L]$ & $E\,[\mathrm{eV}]$ &  $r^{{rms}}_{dd}[\mathrm{pm}]$ &$r^{{rms}}_{d\mu}[\mathrm{pm}]$\\
 \hline
 [0(2),0]& -2988.27  & 0.76 & 0.62\\
 
     [0(2),0]& -2699.04  & 1.57 & 1.16\\
     
     [1,1]& -2889.88  &0.43  &0.49\\
     
     [1,1]& -2665.18  &1.37  &1.06\\

$\mathbf{d\mu(n=1)}$ & -2663.20  & &  & \\

    [0(2),0]& -883.91  & 2.7 & 1.8\\
    
    [0(2),0]& -801.08  &3.4&2.2\\
    
    [0(2),0]& -738.76  & 4.2 &2.7\\
    
    [0(2),0]& -697.59  & 5.7 &3.7\\
    
    [0(2),0]& -678.32  & 8.4 &5.5\\
    
    [0(2),0]& -671.06  & 12.6 &8.4\\
    
    [0(2),0]& -668.05  & 19.0 &13.0\\
    
    [0(2),0]& -666.76  & 28.8 &19.9\\
    
    [0(2),0]& -666.21  &44.3  &30.8\\
    
    [0(2),0]$^{\boldsymbol{\dagger}}$& -665.96  & 69.8 & 48.8\\
    
    [1,0]& -686.96  & 6.6 &4.3\\
    
    [1,0]& -675.21  & 9.5 &6.3\\
    
    [1,0]& -669.88  & 14.2 &9.6\\
    
    [1,0]& -667.56  & 21.3 &14.6\\
    
    [1,0]& -666.56  & 32.4 &22.4\\
    
    [1,0]& -666.12  & 50.1 &34.9\\
    
    [1,0]$^{\boldsymbol{\dagger}}$& -665.92  & 76.3  &53.4\\
    
    [0(2),1]& -688.44  & 3.3 &2.3\\
    
    [0(2),1]& -685.92  & 3.3 &2.5\\
    
    [0(2),1]& -674.60  & 4.8 &3.5\\
    
    [0(2),1]& -669.55  & 7.3 &5.1\\
    
    [0(2),1]& -667.39  & 11.0 &7.7\\
    
    [0(2),1]& -666.47  & 17.2 &12.0\\
    
    [0(2),1]& -666.04  & 29.3  &20.6\\
    
   [1,1]& -877.70  &1.4  &1.3\\
   
    [1,1]& -795.98  & 1.7 &1.5\\
    
    [1,1]& -734.51  & 2.2&1.8 \\  
    
    [1,1]& -694.78  & 2.9 &2.2\\ 
    
    [1,1]& -677.07  & 4.3 &3.1 \\
    
    [1,1]& -670.47  & 6.6 &4.6\\
    
    [1,1]& -667.76  &10.0&7.0\\
    
    [1,1]& -666.62  & 15.6 &10.9\\
    
    [1,1]& -666.09  & 25.1   &17.7\\

 [1,1]$^{\boldsymbol{\dagger}}$& -665.90  & 32.4  &22.9\\
 
 $\mathbf{d\mu(n=2)}$ & -665.80 && & \\
\hline \hline
\end{tabular*}
\end{table}

\begin{table}[htbp]
\renewcommand{\arraystretch}{1.0}
\centering
\caption{\label{tab:ppm} The binding energies and rms radii of  $pp\mu$ are presented below. Newly identified states are marked with a dagger ($\dagger$).}

\begin{tabular*}{\hsize}{@{}@{\extracolsep{\fill}}lcccccc@{}}

\hline\hline
  $[S_{pp},L]$ & $E\,[\mathrm{eV}]$   & $r^{{rms}}_{pp}[\mathrm{pm}]$ &$r^{{rms}}_{p\mu}[\mathrm{pm}]$\\
 \hline 
  [0,0] & -2781.64  & 0.90 & 0.71\\
  
    [1,1]& -2635.76 & 0.56& 0.58\\

   $\mathbf{p\mu(n=1)}$ & -2528.49  &  & \\

  [0,0]& -823.74 & 3.0 & 1.9\\
  
  [0,0]& -725.17 & 4.0 & 2.6\\
  
  [0,0]& -664.08 & 5.8 & 3.7\\
  
  [0,0]& -641.13 & 10.0 & 6.6\\
  
   [0,0]& -634.91 & 17.5 & 11.9\\
   
   [0,0]& -633.00 & 30.8 & 21.3\\
   
  [0,0]$^{\boldsymbol{\dagger}}$& -632.40 & 54.4 & 37.9\\

  [1,0]& -649.68 & 7.4 & 4.8\\
  
   [1,0]& -637.74 & 12.4 & 8.3\\
   
   [1,0]& -633.90 & 21.8 & 14.9\\
   
   [1,0]& -632.68 & 38.4 & 26.7\\
   
   [1,0]& -632.29 & 69.3 & 48.4\\

   [0,1]& -648.13 & 3.8 & 2.9\\
   
   [0,1]& -645.69 & 3.9 & 2.6\\
   
   [0,1]& -636.99 & 6.5 & 4.6\\
   
   [0,1]& -633.60 & 11.7 & 8.2\\
   
   [0,1]& -632.57 & 20.8 & 14.6\\
   
   [0,1]& -632.24 & 36.4 & 25.6\\

   [1,1]& -813.59 & 1.5 & 1.4\\
   
   [1,1]& -717.91 & 2.1 & 1.8\\
   
   [1,1]& -659.91 & 3.0 & 2.3\\
   
   [1,1]& -639.67 & 5.3 & 3.8\\ 
   
   [1,1]& -634.37 & 9.5 & 6.7\\
   
   [1,1]& -632.80 & 17.0 &11.9\\
   
   [1,1]$^{\boldsymbol{\dagger}}$& -632.30 & 32.0 & 22.5\\

  $\mathbf{p\mu(n=2)}$  & -632.12 & & \\
 \hline \hline
\end{tabular*}
\end{table}

\begin{table}[htbp]
\renewcommand{\arraystretch}{1.0}
\centering
\caption{\label{tab:ttm} The binding energies and rms radii of  $tt\mu$  are presented below. }

\begin{tabular*}{\hsize}{@{}@{\extracolsep{\fill}}lccccccc@{}}

\hline\hline
  $[S_{tt},L]$ & $E\,[\mathrm{eV}]$   & $r^{{rms}}_{tt}[\mathrm{pm}]$ &$r^{{rms}}_{t\mu}[\mathrm{pm}]$\\
 \hline
 \hline
  [0,0] & -3074.1  & 0.71 & 0.59 \\
  
     [0,0] & -2795.0  & 1.21 & 0.92 \\
     
     [1,1] & -3000.4  & 0.38 & 0.47 \\
     
   [1,1] & -2756.4  & 0.68 & 0.64 \\
   
     $\mathbf{t\mu(n=1)}$ & -2711.24& &  \\
  
     [0,0] & -907.94  & 2.6 & 1.7 \\
     
     [0,0] & -835.37  & 3.1 & 2.0 \\
     
     [0,0] & -776.83  & 3.8 & 2.5 \\
     
     [0,0] & -732.33  & 4.8 & 3.1 \\
     
     [0,0] & -703.81  & 6.0 & 3.9 \\
     
     [0,0] & -689.92  & 8.1 & 5.3 \\
     
     [0,0] & -683.70  & 11.5 & 7.6 \\
     
     [0,0] & -680.71  & 16.2 & 11.0 \\
     
     [0,0] & -679.24  & 23.0 & 15.8 \\
     
     [0,0] & -678.52  & 31.1 & 21.5 \\
     
     [0,0] & -678.17  & 45.4 & 31.6  \\
     
     [0,0] & -677.95  & 75.2 & 52.6 \\

   [1,0] & -700.74  & 6.3 & 4.1 \\
   
   [1,0] & -689.70  & 8.4 & 5.6 \\
   
   [1,0] & -683.78  & 11.7 & 7.8 \\
   
   [1,0] & -680.78  & 16.4 & 11.2 \\
   
 [1,0] & -679.29  & 23.1 & 15.9 \\
 
   [1,0] & -678.56  & 33.6 & 23.3 \\
   
   [1,0] & -678.17  & 50.7 & 35.3 \\
   
   [1,0] & -677.96  & 79.5 & 55.7 \\

   [0,1] & -705.24  & 3.1 & 2.1 \\
   
   [0,1] & -699.98  & 3.2 & 2.4 \\
   
   [0,1] & -689.16  & 4.3 & 3.1 \\
   
   [0,1] & -683.43  & 6.0 & 4.2 \\

   [0,1] & -680.55  & 8.6 & 6.0 \\
   
   [0,1] & -679.13  & 12.5 & 8.7 \\
   
   [0,1] & -678.43  & 19.0 & 13.3 \\
   
   [0,1] & -678.01  & 32.5 & 23.0 \\

     [1,1] & -903.45  & 1.3 & 1.3 \\
     
     [1,1] & -831.25  & 1.6 & 1.4 \\

     [1,1] & -773.07  & 1.9 & 1.6 \\
     
     [1,1] & -729.03  & 2.4 & 1.9 \\

     [1,1] & -700.85  & 3.2 & 2.4 \\
     
     [1,1] & -688.17  & 4.6 & 3.3 \\
     
     [1,1] & -682.73  & 6.5 & 4.6 \\
     
     [1,1] & -680.18  & 9.3 & 6.5 \\
     
     [1,1] & -678.93  & 13.4 & 9.3 \\
     
     [1,1] & -678.34  & 19.5 & 13.6 \\
     
     [1,1] & -677.98  & 32.6 & 23.1 \\

    $\mathbf{t\mu(n=2)}$ &-677.81  & &\\
\hline \hline
\end{tabular*}
\end{table}

Using the ESVM, we obtain new shallow states below the $d\mu(n=2)$ threshold in $dd\mu$ system and below the $t\mu(n=2)$ threshold in $dt\mu$ system.
We also identify shallow resonances lying between the nearly degenerate $t\mu(n=2)$ and $d\mu(n=2)$ thresholds, as shown in Fig. \ref{fig:dtm}.
These newly identified states are marked with a dagger in Tables \ref{tab:dtm} and \ref{tab:ddm}.
The internal structure of the resonant states lying between the $t\mu(n=2)$ and $d\mu(n=2)$ thresholds can be inferred from their rms radii.
As the resonance energies evolve from the vicinity of the $t\mu(n=2)$ threshold toward the $d\mu(n=2)$ threshold, the rms distance $r^{\mathrm{rms}}_{t\mu}$ generally increases, while $r^{\mathrm{rms}}_{d\mu}$ decreases.
This behavior indicates a gradual structural rearrangement, with these states becoming increasingly similar to quasi-bound configurations of the $d\mu(n=2)+t$ channel.

The $pp\mu$ system is structurally similar to $dd\mu$, differing only by the mass of the identical nuclear pair.
As shown in Table~\ref{tab:ppm}, we identify two additional shallow states below the $p\mu(n=2)$ threshold in the $L=0$ and $L=1$ sectors, compared to previous investigations~\cite{lindroth2003decay,kilic2004coulombic}.

Although the bound states of the $tt\mu$, $pd\mu$, and $pt\mu$ molecular ions have been extensively investigated in previous studies~\cite{Frolov:2011yfh}, their resonant states have received little attention and have not been systematically studied.
In this work, we present a complete set of bound and resonant states lying below the corresponding $X\mu(n=2)$ thresholds.
The full complex-energy spectra obtained in our calculations for the $tt\mu$, $pd\mu$, and $pt\mu$ systems are summarized in Tables~\ref{tab:ttm}, \ref{tab:pdm}, and \ref{tab:ptm}, respectively.
These results include comprehensive sets of resonant states below the $n=2$ thresholds, which have not been reported in previous works.

Among all hydrogen-molecule–like muonic ions, only the $dt\mu$, $pd\mu$, and $pt\mu$ systems exhibit resonant states with visible decay widths of order $\mathcal{O}(10^{-1})~\mathrm{eV}$.
These broader widths arise from the strong coupling to the nearly degenerate $n=2$ atomic thresholds.
In contrast, the remaining systems possess extremely narrow resonances, making them suitable candidates for treatments based on the Born–Oppenheimer approximation.

\begin{table}[htbp]
\renewcommand{\arraystretch}{1.0}
\centering
\caption{\label{tab:pdm} The binding energies and rms radii of  $pd\mu$ are presented below. }

\begin{tabular*}{\hsize}{@{}@{\extracolsep{\fill}}lcccccc@{}}

\hline\hline
  $L$ & $E\,[\mathrm{eV}]$   & $r^{{rms}}_{pd}[\mathrm{pm}]$ &$r^{{rms}}_{p\mu}[\mathrm{pm}]$&$r^{{rms}}_{d\mu}[\mathrm{pm}]$\\
 \hline
  0& -2884.75 & 0.84 & 0.73 & 0.62 \\
   1& -2760.70 & 0.51 & 0.58 & 0.50 \\

 $\mathbf{d\mu(n=1)}$ & -2663.20  & &  & \\

   $\mathbf{p\mu(n=1)}$ & -2528.49  &  & \\

   0& -853.18  & 2.8 & 1.9 & 1.8 \\
   0& -761.14  & 3.7 & 2.5 & 2.2 \\
   0& -700.61  & 5.2 & 4.2 & 2.3 \\
   0& -678.11  & 8.3 & 7.5 & 1.9 \\
   0& -670.19  & 13.6 & 12.8 & 1.8 \\
   0& -667.37  & 22.5 & 21.8 & 1.7 \\
   0& -666.36  & 37.3 & 36.5 & 1.7 \\
   0& -666.00  & 57.7 & 57.0 & 1.7 \\
   1& -844.84  & 1.4 & 1.4 & 1.4 \\
   1& -754.92  & 1.9 & 1.7 & 1.6 \\
   1& -696.84  & 2.7 & 2.5 & 1.6 \\
   1& -676.51  & 4.4 & 4.2 & 1.6 \\
   1& -669.57  & 4.1 & 3.5 & 1.9 \\
   1& -669.52  & 7.2 & 7.1 & 1.5 \\
  1& -667.09  & 12.3 & 12.1 & 1.5 \\
   1& -666.25  & 19.9 & 19.7 & 1.5 \\
 $\mathbf{d\mu(n=2)}$ & -665.80 && & \\
   0& -640.77 - 0.5i & 9.8 & $4.1^*$ & 9.4 \\
   0& -635.31 - 0.2i & 16.4 & 2.8 & 15.6 \\
 0& -633.31 - 0.1i & 26.5 & 2.4 & 25.7 \\
   0& -632.57   & 41.7 & 2.1 & 40.9 \\
   1& -639.97 - 0.6i & 10.3 & 9.1 & 5.0 \\
   1& -634.96 - 0.2i & 7.6 & $1.7^*$ & 8.3 \\
   1& -633.15 - 0.1i & 14.1 & 2.2 & 13.8 \\
   1& -632.50  & 21.2 & 2.0 & 21.1 \\

  $\mathbf{p\mu(n=2)}$  & -632.12 & & & \\
\hline \hline
\end{tabular*}
\end{table}

\begin{table}[htbp]
\renewcommand{\arraystretch}{1.0}
\centering
\caption{\label{tab:ptm} The binding energies and rms radii of  $pt\mu$ are presented below. }

\begin{tabular*}{\hsize}{@{}@{\extracolsep{\fill}}lcccccc@{}}

\hline\hline
  $L$ & $E\,[\mathrm{eV}]$   & $r^{{rms}}_{pt}[\mathrm{pm}]$ &$r^{{rms}}_{p\mu}[\mathrm{pm}]$&$r^{{rms}}_{t\mu}[\mathrm{pm}]$\\
 \hline
     0& -2925.08  & 0.82 & 0.73 & 0.59 \\
   1 & -2810.37  & 0.49 & 0.57 & 0.49 \\

    $\mathbf{t\mu(n=1)}$ & -2711.24& & & &\\

   $\mathbf{p\mu(n=1)}$ & -2528.49  &  & & \\

  0& -864.22  & 2.8 & 1.9 & 1.8 \\
   0& -775.28  & 3.6 & 2.5 & 2.1 \\
   0& -715.51  & 5.0 & 4.0 & 2.2 \\
   0& -691.44  & 7.9 & 7.1 & 1.8 \\
   0& -682.82  & 12.7 & 12.0 & 1.7 \\
   0& -679.67  & 20.6 & 19.9 & 1.6 \\
   0& -678.50  & 33.2 & 32.5 & 1.6 \\
   0& -678.07  & 53.7 & 52.9 & 1.6 \\
   1 & -856.59  & 1.4 & 1.4 & 1.3 \\
   1 & -769.49  & 1.8 &  1.7 & 1.5 \\
   1 & -711.87  & 2.6 &  2.4 & 1.6 \\ 
  1 & -689.87  & 4.1 &  4.0 & 1.5 \\
     1 & -682.15  & 6.7 &  6.5 & 1.5 \\
     1 & -679.46  & 4.5 &  4.1 & 1.7 \\
     1 & -679.38  & 11.0 &  10.8 & 1.5 \\
     1 & -678.38  & 17.5 &  17.3 & 1.5 \\
     1 & -678.01  & 32.8 &  32.9 & 1.6 \\

   $\mathbf{t\mu(n=2)}$ &-677.81 & & &\\
0& -659.73 - 0.2i & 6.2 & 3.5 & 5.2 \\
  0& -642.80 - 0.1i & 9.4 & 2.8 & 8.4 \\
   0& -636.31 - 0.1i & 14.3 & 1.0 & 13.6 \\
   0& -633.78  & 22.6 & 1.7 & 21.8 \\
   0& -632.78  & 36.0 & 1.7 & 35.2 \\
   0& -632.38  & 60.5 & 1.8 & 59.7 \\
     1 & -658.06 - 0.3i & 3.7 &   2.9 & 3.0 \\
     1 & -641.91 - 0.2i & 5.1 &  2.5 & 4.6 \\
     1 & -635.89 - 0.1i & 7.4 &   1.4 & 7.2 \\
     1 & -633.59  & 11.9 &   1.6 & 11.7 \\
     1 & -632.69  & 18.9 &   1.6 & 18.7 \\
        1 & -632.34  & 30.6 &   1.6 & 30.3 \\
   $\mathbf{p\mu(n=2)}$  & -632.12 & & & \\
\hline \hline
\end{tabular*}
\end{table}

\section{Four-body systems}~\label{sec:4body}

\begin{figure}[htbp]
  \centering
  \includegraphics[width=0.45\textwidth]{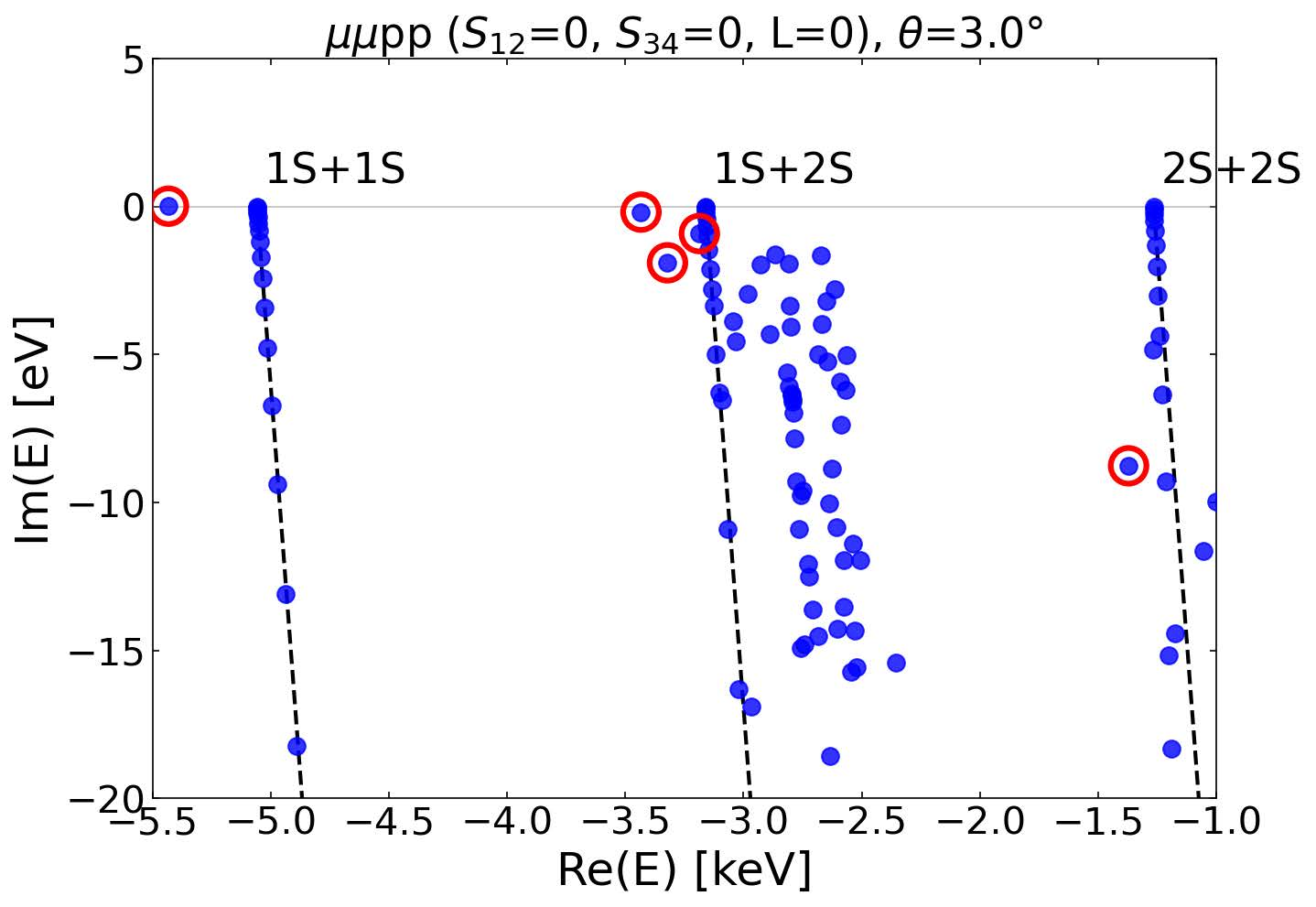} 
  \caption{\label{fig:mmpp} Complex-energy spectra of the $\mu \mu pp$ system ($S_{12}=0$, $S_{34}=0$, $L=0$, $\theta=6^\circ$) . }
    \setlength{\belowdisplayskip}{1pt}
\end{figure}

\begin{table*}[htbp]
\renewcommand{\arraystretch}{1.0}
\centering
\caption{\label{tab:mmpp} Complex eigenenergies $E$ of the $\mu\mu pp$ system.
Benchmark results from previous calculations are listed in the last column for comparison.
The column “Type” indicates bound states (“B”) and resonant states (“R”) in the pure Coulomb system.}

\begin{tabular*}{\hsize}{@{}@{\extracolsep{\fill}}lcccccc@{}}

\hline\hline
$[S_{\mu\mu},S_{pp},L]$ &    $E\,[\mathrm{eV}]$ &  Type &$r^{{rms}}_{p\mu}[\mathrm{pm}]$ &  $r^{{rms}}_{pp}[\mathrm{pm}]$  & $r^{{rms}}_{\mu\mu}[\mathrm{pm}]$ &$E_{bench}\,[\mathrm{eV}]$\\
 \hline
 [0,0,0] &  -5431.91  &B  & 0.62 & 0.64 & 0.81 &-5431.5~\cite{Hamahata:2001skx}\\

[0,1,1] &  -5147.57  &B  & 0.56 & 0.44 & 0.77 &-5147.3~\cite{Hamahata:2001skx}\\

  $\mathbf{p\mu(n=1)+p\mu(n=1)}$&  -5056.99 & &\\
 
[0,0,0] &    -3433.18-0.4i   &R & 1.4 & 1.5 & 1.5 &-3370-0.16i~\cite{Yang:2025oih}\\

 [0,0,0] &   -3321.23-2.5i &R   & 1.6 & 1.2 & 2.1 & -3259-4.3i~\cite{Yang:2025oih}\\
 
 [0,0,0] &   -3186.55-0.9i &R & 2.0 & 2.6 & 2.1 & \\

   [1,0,0] &   -3430.15 &B & 1.4 & 0.8 & 1.9 & -3406~\cite{Yang:2025oih}\\

 [0,1,0] &   -3554.44 &B & 1.1 & 1.2 & 1.2 & -3449~\cite{Yang:2025oih}\\

  [0,1,0] &   -3296.14 &B & 1.5 & 1.9 & 1.4 & -3194~\cite{Yang:2025oih}\\

  [0,0,1] &   -3523.22  &B & 0.8 & 0.7 & 1.0 & \\

  [0,0,1] &   -3317.24 &B & 1.0 & 0.9 & 1.1 & \\
 
  [0,0,1] &   -3184.78  &B & 1.3 & 1.4 & 1.7 & \\

 [1,0,1] &   -3345.64-26.9i &R & 1.0 & 1.1 & 1.2 & \\
  
  [0,1,1] &   -3385.60-1.3i &R & 1.0 & 0.8 & 1.3 & \\

  [0,1,1] &   -3191.26-1.7i &R & 1.4 & 1.0 & 1.9 & \\

  [1,1,1] &   -3254.64 &B & 1.4 & 0.5 & 1.9 & \\

 $\mathbf{p\mu(n=1)+p\mu(n=2)}$ & -3160.62 & & &\\

 [0,0,0] &  -1374.56-10.3i &R & 4.0 & 5.1 & 5.2 & \\
 
 [0,1,1] &   -1347.06-13.2i  &R & 2.1 & 2.0 & 3.0 & \\
 $\mathbf{p\mu(n=2)+p\mu(n=2)}$ & -1264.25 & \\

\hline \hline
\end{tabular*}
\end{table*}

Compared with three-body systems, four-body muonic ions contain a much richer set of dissociation subchannels, including both atomic fragments and muonic molecular-ion configurations.
This complexity makes them an excellent testing ground for the performance of the ESVM, particularly in resolving near-threshold resonances embedded in complicated continua.

Figure.~\ref{fig:mmpp} shows a representative complex-energy spectrum of the S-wave $\mu\mu pp$ system with $S_{12}=0$ and $S_{34}=0$, obtained from a calculation in which the scattering-like basis functions are constructed using $p\mu(n\leq2)$ configurations.
In this illustrative calculation, the parameters of the scattering basis are optimized by minimizing the energy of the $p\mu(2S)$ subsystem.
As a result, the $p\mu(1S)+p\mu(1S)$ and $p\mu(1S)+p\mu(2S)$ continua are cleanly rotated into straight rays in the complex-energy plane, indicating that these channels are properly reconstructed.

In contrast, scattering states associated with higher atomic excitations, such as $p\mu(n=1)+p\mu(n=m)$ with $m>2$, as well as three-body breakup channels, are not explicitly optimized in this illustrative basis and therefore appear irregularly distributed and poorly resolved above the $p\mu(n=1)+p\mu(n=2)$ threshold.

Nevertheless, because the basis is optimized for thresholds up to $n=2$, the $p\mu(2S)+p\mu(2S)$ scattering channel is clearly constructed.
Even in the presence of these unresolved background states, the ESVM succeeds in extracting the physically relevant near-threshold resonances associated with the $p\mu(2S)+p\mu(2S)$ channel.
This behavior demonstrates that the ESVM can reliably isolate resonant states governed by a given threshold, even when higher-lying scattering channels are not explicitly controlled.

The bound and resonant states obtained for the $\mu\mu pp$ system are summarized in Table~\ref{tab:mmpp}, together with benchmark values from Gaussian expansion method (GEM) calculations~\cite{Yang:2025oih,Hamahata:2001skx} for comparison.
Our results systematically yield deeper energies for the resonant states, while the bound-state energies are only slightly deeper than the benchmark values.
In addition, our calculations reveal a number of resonant states, including several located in the vicinity of the $p\mu(2S)+p\mu(2S)$ threshold, as well as other resonances that have not been systematically resolved in previous studies. These states are classified accordingly as bound (“B”) or resonant (“R”) in the tables. The lowest channel with quantum numbers $[S_{\mu\mu}, S_{pp}, L]=[0,0,0]$ is strongly attractive and supports a deeply bound state.
In the infinite proton-mass limit, the channel with quantum numbers
$[S_{\mu\mu}, S_{pp}, L]=[0,1,1]$
becomes degenerate with the $[0,0,0]$ channel, giving rise to a corresponding deeply bound state below the $p\mu(1S)+p\mu(1S)$ threshold.

It is also noteworthy that some bound states appear at energies above the $p\mu(1S)+p\mu(1S)$ dissociation threshold.
This behavior is a direct consequence of the Pauli principle and exchange symmetry for identical particles: states with quantum numbers
$[S_{\mu\mu},S_{pp},L]=[0,1,0]$, $[1,0,0]$, $[0,0,1]$, and $[1,1,1]$
cannot be constructed from two ground-state muonic hydrogen atoms $p\mu(1S)+p\mu(1S)$.
As a result, these states do not couple to the lowest atomic dissociation channel and therefore remain bound in the pure Coulomb system, even though their energies lie above the $p\mu(1S)+p\mu(1S)$ threshold.

For completeness, the bound and resonant states of the $\mu\mu dd$ and $\mu\mu tt$ systems lying below the second dissociation thresholds
$$d\mu(1S)+d\mu(2S), \qquad t\mu(1S)+t\mu(2S)$$
are summarized in Tables~\ref{tab:mmdd} and~\ref{tab:mmtt}, respectively.

\begin{table*}[htbp]
\renewcommand{\arraystretch}{1.0}
\centering
\caption{\label{tab:mmdd}  Complex eigenenergies $E$ of the $\mu\mu dd$ system.
Benchmark results from previous calculations are listed in the last column for comparison.
The column “Type” indicates bound states (“B”) and resonant states (“R”) in the pure Coulomb system.}

\begin{tabular*}{\hsize}{@{}@{\extracolsep{\fill}}lcccccc@{}}

\hline\hline
$[S_{\mu\mu},S_{dd},L]$ &   $E\,[\mathrm{eV}]$ &  Type &$r^{{rms}}_{d\mu}[\mathrm{pm}]$ &  $r^{{rms}}_{dd}[\mathrm{pm}]$  & $r^{{rms}}_{\mu\mu}[\mathrm{pm}]$ &$E_{bench}\,[\mathrm{eV}]$\\
 \hline
 [0,0(2),0] &  -5831.92   &B & 0.55 & 0.54 & 0.73 &-5831.6~\cite{Hamahata:2001skx}\\

 [0,0(2),0] &  -5331.16  &B  & 1.54 & 2.09 & 2.17 &-5328.6~\cite{Hamahata:2001skx}\\

 [0,1,1] &  -5633.48  &B  & 0.49 & 0.31 & 0.68 &-5634.2~\cite{Hamahata:2001skx}\\
 
  $\mathbf{d\mu(n=1)+d\mu(n=1)}$  & -5326.40 & &\\
 
[0,0(2),0] &    -3664.89-0.5i  &R & 1.2 & 1.3 & 1.4 &\\

 [0,0(2),0] &   -3572.28-0.2i  &R  & 1.5 & 1.0 & 1.9 & \\
 
 [0,0(2),0] &   -3431.78-0.2i  &R & 1.6 & 2.0 & 1.5 & \\

[1,0(2),0] &   -3675.3  &B & 1.4 & 0.8 & 1.9 & \\

[1,0(2),0] &   -3344.4 &B  & 1.4 & 0.8 & 1.9 & \\
   
   [0,1,0] &   -3823.08 &B  & 1.0 & 1.0 & 1.1 & \\

 [0,1,0] &   -3591.49 &B & 1.2 & 1.5 & 1.2 & \\

 [0,1,0] &   -3414.02 &B & 1.6 & 2.1 & 1.5 & \\
 
  [1,1,0] &   -3359.55-1.8i &R & 1.8 & 1.1 & 2.5 & \\
  
  [0,0(2),1] &   -3784.51 &B & 0.7 & 0.6 & 0.94 & \\

   [0,0(2),1] &   -3585.46 &B & 0.9 & 0.8 & 1.0 & \\

    [0,0(2),1] &   -3457.85 &B  & 1.0 & 1.0 & 1.3 & \\

    [0,0(2),1] &   -3344.95 &B  & 1.3 & 1.4 & 1.6 & \\

[1,0(2),1] &   -3607.32-17.1i &R & 1.1 & 0.9 & 0.9 & \\

    [0,1,1] &   -3630.54-1.3i &R & 1.0 & 0.7 & 1.2 & \\

    [0,1,1] &   -3477.23-1.1i &R & 1.4 & 0.6 & 1.9 & \\

    [0,1,1] &   -3386.86-3.3i  &R & 1.2 & 1.1 & 1.5 & \\

    [1,1,1] &   -3556.66  &B & 1.3 & 0.4 & 1.8 & \\

 $\mathbf{d\mu(n=1)+d\mu(n=2)}$ &  -3329.00&  & &\\

\hline \hline
\end{tabular*}
\end{table*}

\begin{table*}[htbp]
\renewcommand{\arraystretch}{1.0}
\centering
\caption{\label{tab:mmtt}  Complex eigenenergies $E$ of the $\mu\mu tt$ system.
Benchmark results from previous calculations are listed in the last column for comparison.
The column “Type” indicates bound states (“B”) and resonant states (“R”) in the pure Coulomb system.}

\begin{tabular*}{\hsize}{@{}@{\extracolsep{\fill}}lcccccc@{}}

\hline\hline
$[S_{\mu\mu},S_{tt},L]$ &   $E\,[\mathrm{eV}]$   &Type &$r^{{rms}}_{t\mu}[\mathrm{pm}]$ &  $r^{{rms}}_{tt}[\mathrm{pm}]$  & $r^{{rms}}_{\mu\mu}[\mathrm{pm}]$ &$E_{bench}\,[\mathrm{eV}]$\\
 \hline
 [0,0,0] &  -5998.08  &B  & 0.53 & 0.50 & 0.70 &-5998.0~\cite{Hamahata:2001skx}\\

 [0,0,0] &  -5486.18   &B & 0.80 & 0.96 & 1.09 &-5486.1~\cite{Hamahata:2001skx}\\

 [0,1,1] &  -5847.34  &B  & 0.46 & 0.27 & 0.65 &-5848.6~\cite{Hamahata:2001skx}\\

 [0,1,1] &  -5425.66  &B  & 0.62 & 0.64 & 0.81 &-5425.2~\cite{Hamahata:2001skx}\\
 
  $\mathbf{t\mu(n=1)+t\mu(n=1)}$  & -5422.48&\\
 
[0,0,0] &    -3754.15 -0.4i &R  & 1.2 & 1.3 & 1.3&\\

 [0,0,0] &   -3675.90-0.4i  &R  & 1.4 & 0.9 & 1.9 & \\
 
 [0,0,0] &   -3537.42-1.6i &R & 1.5 & 1.8 & 1.4 & \\

 [0,0,0] &   -3402.12-4.2i &R & 2.0 & 2.7 & 2.2 & \\

[1,0,0] &   -3776.56  &B & 1.3   & 0.7 & 1.8 & \\

[1,0,0] &   -3451.76 &B & 1.6 & 1.1 & 2.3 & \\
   
 [0,1,0] &   -3930.60 &B & 1.0 & 0.9 & 1.1 & \\

 [0,1,0] &   -3719.82 &B & 1.2 & 1.3 & 1.1 & \\

  [0,1,0] &   -3537.17 &B & 1.5 & 1.8 & 1.3 & \\

   [0,1,0] &   -3409.91 &B & 1.9 & 2.3 & 2.2 & \\

   [1,1,0] &   -3463.03-1.8i &R & 1.7 &   0.9 & 2.4\\

    [0,0,1] &   -3891.64 &B & 0.7 & 0.5 & 0.9 & \\

    [0,0,1] &   -3698.57 &B & 0.8 & 0.7 & 1.0 & \\

    [0,0,1] &   -3581.05 &B & 1.0 & 0.9 & 1.2 & \\

    [0,0,1] &   -3453.70 &B & 1.1 & 1.1 & 1.4 & \\

   [1,0,1] &   -3718.12-2.4i &R & 0.8 &   0.9 & 1.0\\

  [1,0,1] &   -3424.10-1.6i &R  &1.3 & 0.9   & 1.8\\

    [1,0,1] &   -3408.64-2.2i &R & 1.3 &   1.6 & 1.6\\
   
    [0,1,1] &   -3725.17-0.1i &R & 0.9 & 0.7 & 1.2 & \\

    [0,1,1] &   -3600.93-0.9i &R & 1.3 & 0.5 & 1.9 & \\

    [0,1,1] &   -3496.08-2.8i &R & 1.1 & 0.9 & 1.4 & \\

    [1,1,1] &   -3685.95 &B & 1.2 & 0.4 & 1.8 & \\

    [1,1,1] &   -3400.45 &B & 1.4 & 0.8 & 1.9 & \\

 $\mathbf{t\mu(n=1)+t\mu(n=2)}$ &-3389.05 & &\\

\hline \hline
\end{tabular*}
\end{table*}

\section{Summary and Discussion}~\label{sec:sum}

We have investigated the bound and resonant states of a broad class of few-body muonic ion systems, including hydrogen-like muonic ions ($\mu\mu X$, $X=p,d,t$) and muonic molecular ions ($pp\mu$, $dd\mu$, $tt\mu$, $pd\mu$, $pt\mu$, $dt\mu$), and the four-body double-muonic hydrogen molecules ($\mu\mu pp,\mu\mu dd,\mu\mu tt$).
The calculations were performed using an extended stochastic variational method (ESVM) combined with the complex scaling method, allowing bound and quasibound states to be treated on the same footing.

For the hydrogen-like systems, our fully dynamical three-body calculations reproduce the benchmark energies of the ground and lowest resonant states with discrepancies below $0.01~\mathrm{eV}$, and they provide a complete mapping of all resonances below the $X\mu(n=2)$ thresholds.
The spatial structure of these states, characterized through rms distances, shows clear correlations between binding strength and compactness, while the imaginary parts of the rms radii remain negligible for narrow resonances.

For the muonic molecular ions relevant to $\mu$CF, our results recover the well-known shallow $S_{12}=1$, $L=1$ bound states in $dt\mu$ and $dd\mu$, and provide additional shallow levels near the $n=2$ thresholds, including states not reported in previous studies.
The dense accumulation of resonances between the nearly degenerate $d\mu(2S)$ and $t\mu(2S)$ thresholds is reproduced, and several new near-threshold states are identified.
For systems with unequal nuclei ($pd\mu$, $pt\mu$), we obtain the complete complex-energy spectra below the corresponding $n=2$ thresholds.

The four-body double-muonic systems provide a stringent test of the ESVM due to the coexistence of atomic, molecular, and three-body breakup subchannels.
For the $\mu\mu pp$ system, the method correctly reconstructs the $p\mu(1S)+p\mu(1S)$, $p\mu(1S)+p\mu(2S)$, and $p\mu(2S)+p\mu(2S)$ continua, and resolves stable near-threshold resonances embedded in a dense background.
Beyond reproducing known bound states, our calculations systematically yield deeper energies for the resonant states, while the bound-state energies are only slightly deeper than benchmark values.
Moreover, a large number of previously unexplored resonant states are identified.

Among all molecular ions studied, only $dt\mu$, $pd\mu$, and $pt\mu$ exhibit resonances with widths at the level of $0.1$–$1~\mathrm{eV}$, reflecting enhanced channel coupling near nearly degenerate thresholds; all other resonances are extremely narrow and thus suitable for Born–Oppenheimer-type treatments.

Overall, the ESVM–CSM framework developed here provides a robust tool for resolving near-threshold structures in few-body muonic systems with high precision.
The spectra obtained in this work supply an unified reference for future studies of muonic atoms, resonant molecular formation in $\mu$CF, and electromagnetic processes involving exotic Coulombic systems.

\begin{acknowledgements}

We are grateful to Yao Ma, Meng Lu, Yan-Ke Chen, Hui-Ming Yang and Wei-Lin Wu for the helpful discussions. 
This project was supported by the National Natural Science Foundation of China (Grant No. 12475137). The computational resources were supported by High-performance Computing Platform of Peking University.

\end{acknowledgements}

\bibliography{Ref}

\end{document}